\documentclass[11pt]{amsart}

\usepackage{amscd}
\usepackage{amsmath}
\usepackage{graphicx}
\usepackage{amsfonts}
\usepackage{amssymb}
\usepackage{curves,epic}
\usepackage{caption}
\begin{document}
\title[Indecomposability of entanglement witnesses]
{Indecomposability of entanglement witnesses constructed from any
permutations}
\author{Xiaofei Qi}
\address[Xiaofei Qi]{
Department of Mathematics, Shanxi University, Taiyuan 030006, P. R.
China} \email{xiaofeiqisxu@aliyun.com}

\author{Jinchuan Hou}
\address[Jinchuan Hou]{Department of
Mathematics, Taiyuan University of Technology, Taiyuan 030024, P. R.
 China} \email{jinchuanhou@aliyun.com}

\thanks{ This work is partially supported by National Natural Science Foundation of
China (11171249,11101250) and  Youth Foundation of  Shanxi Province
(2012021004).}

\begin{abstract}

Let $n\geq 2$ and $\Phi_{n,t,\pi}: M_n({\mathbb C}) \rightarrow
M_n({\mathbb C})$ be a linear map defined by
$\Phi_{n,t,\pi}(A)=(n-t)\sum_{i=1}^nE_{ii}AE_{ii}+t\sum_{i=1}^nE_{i,\pi(i)}AE_{i,\pi(i)}^\dag-A$,
where   $0\leq t\leq n$, $E_{ij}$s are the matrix units and $\pi$ is
a non-identity permutation of $(1,2,\cdots,n)$. Denote by $\{{ F}_s:
s=1,2\ldots, k\}$ the set of all minimal cycles of $\pi$ and
$l(\pi)=\max\{\# { F}_s: s=1,2,\ldots,k\}$ the length of $\pi$. It
is shown that the Hermitian matrix $W_{n,t,\pi}$   induced by
$\Phi_{n,t,\pi}$ is an indecomposable entanglement witness if and
only if $\pi^2\not={\rm id}$ (the identity permutation) and
$0<t\leq\frac{n}{l(\pi)}$. Some new bounded entangled states are
detected by such witnesses that cannot be distinguished by PPT
criterion, realignment criterion, etc..

\end{abstract}

\maketitle
\section{Introduction}

Entanglement is an important physical resource to realize various
quantum information and quantum communication tasks such as
teleportation, dense coding, quantum cryptography and key
distribution \cite{LK,NC}. One of the most important topics in the
theory of entanglement is how to distinguish entangled states from
separable states. Entanglement witnesses provide one of the best
known methods of entanglement detection in bipartite composite
quantum systems  (see \cite{Hor}).

\if  Recall that a quantum state on $H$ is a density operator $\rho$
which is positive semi-definite and has trace 1. Denote by
${\mathcal S}(H)$ the set of all states on $H$ and $ {\mathcal
B}(H)$ the algebra of all bounded linear operators acting on $H$. A
state $\rho\in{\mathcal S}(H\otimes K)$ of the bipartite composition
system $H\otimes K$ is said to be separable if $\rho$ can be written
as $\rho=\sum_{i=1}^k p_i \rho_i\otimes \sigma _i,$ where $\rho_i$
and $\sigma_i$ are states on $H$ and $K$ respectively, and $p_i$ are
positive numbers with $\sum _{i=1}^kp_i=1$. Otherwise, $\rho$ is
entangled.\fi

Let $H$ and $K$ be finite dimensional complex  Hilbert spaces. A
Hermitian matirx $W\in{\mathcal B}(H\otimes K)$
 is an entanglement witness (briefly, EW) if
$W$ is not positive and ${\rm Tr}(W\sigma)\geq 0$ holds for all
separable states $\sigma\in{\mathcal S}(H\otimes K)$.  It is
well-known that if $\rho$ is an entangled state, then there is some
EW $W$ such that ${\rm Tr}(W\rho)<0$ (that is, the entanglement in
$\rho$ can be detected by $W$) \cite{Hor}.   However there is no
universal EW $W$ so that every entangled state can be detected by
$W$. Therefore constructing as many as possible EWs is important to
detect entanglement in states. There was a considerable effort in
constructing and analyzing the structure of EWs \cite{B,CK, HQ,JB,
TG}.

Due to the Choi-Jamio{\l}kowski isomorphism \cite{C,J}, a Hermitian
matrix $W\in{\mathcal B}(H\otimes K)$  is an EW if and only if there
exists a positive linear map which is not completely positive (NCP,
briefly) $\Phi:{\mathcal B}(H)\rightarrow{\mathcal B}(K)$ and a
maximally entangled state $P^+\in{\mathcal B}(H\otimes H)$ such that
$W=W_\Phi=(I_n\otimes \Phi)P^+$. Recall that a maximally entangled
state is a pure state  $P^+=|\psi^+\rangle\langle\psi^+|$ with
$|\psi^+\rangle=\frac{1}{\sqrt{n}}(|11\rangle+|22\rangle+\cdots
|nn\rangle) $, where $n=\dim H$ and $\{|i\rangle\}_{i=1}^n$ is an
orthonormal basis of $H$. Thus, up to a multiple by positive scalar,
$W_\Phi$ can be written as the matrix
$W_\Phi=(\Phi(E_{ij}))_{n\times n}$, where  $E_{ij}=|i\rangle\langle
j|$. For a positive linear map $\Phi:{\mathcal
B}(H)\rightarrow{\mathcal B}(K)$, we always denote $W_\Phi$  the
Choi-Jamio{\l}kowski matrix of $\Phi$ with respect to a given basis
of $H$, that is $W_\Phi=(\Phi(E_{ij}))_{n\times n}$, and we say that
$W_\Phi$ is the EW induced by the NCP positive map $\Phi$.
Conversely, for an EW $W$, we denote by $\Phi_W$   the associated
NCP positive map so that $W=W_{\Phi_W}$.

Recall that an EW $W$ is called decomposable if $W
=Q_1+Q_2^{\Gamma}$ for some operators $Q_1,Q_2\geq0$, where
$Q_2^{\Gamma}$ stands for any one of $Q_2^{T_1}$ and $Q_2^{T_2}$,
 the partial transpose of $Q_2$ with respect to the
subsystems $H$ and $K$, respectively. Otherwise, $W$ is called
indecomposable. Similarly, a positive map $\Delta$ is said to be
decomposable if it is the sum of a completely positive map
$\Delta_1$ and the composition of a completely positive map
$\Delta_2$ and the transpose $\bf T$, i.e.,
$\Delta=\Delta_1+\Delta_2\circ{\bf T}$. It is clear that $W_\Phi$ is
decomposable if and only if $\Phi$ is decomposable. Note  that
decomposable EWs cannot detect PPT (positive partial transpose)
entangled states and, therefore, such EWs are useless in search of
bound entangled state. Unfortunately, there is no general method to
construct indecomposable EWs and only very few examples of
indecomposable EWs are available in the literature
\cite{CK2,Ha2,Ha3}. In this paper we develop a way to construct
indecomposable entanglement witnesses from any permutation $\pi$
with $\pi^2\not=$id.

Let $\pi$ be a permutation of $(1,2,\ldots,n)$. For a subset $F$  of
$\{1,2,\ldots,n\}$, if $\pi(F)=F$, we say $F$ is an invariant subset
of $\pi$.  Let  $F$ be an invariant subset of $\pi$. If $G\subseteq
F$ and $G$ is invariant under $\pi$ imply $G=F$, we say $F$ is a
minimal invariant subset of $\pi$. It is obvious that a minimal
invariant subset is a loop of $\pi$ and $\{1,2,\ldots,
n\}=\cup_{s=1}^k F_s$, where $\{F_s\}_{s=1}^k$ is the set of all
disjoint minimal invariant subsets of $\pi$. Denote by $\# F_s$ the
cardinal number of $F_s$. Then $\sum_{s=1}^k \# F_s=n$.
$l(\pi)=\max\{\# F_s: s=1,2,\ldots,r\}$ is called the length of
$\pi$. In the case that $l(\pi)=n$, $\pi$ is called cyclic. So every
permutation $\pi$ of $(1,\dots, n)$ has a disjoint cyclic
decomposition $\pi=(\pi_1)(\pi_2)\cdots(\pi_k)$, that is, there
exists a set $\{F_s\}_{s=1}^k$ of disjoint cycles of $\pi$ with
$\cup_{s=1}^k F_s=\{1,2,\ldots , n\}$ such that $\pi_s=\pi|_{F_s}$
and $\pi(i)=\pi_s(i)$ whenever $i\in F_s$.

If $\dim H=n$, by fixing an orthonormal basis, one may identify
${\mathcal B}(H)$ with $M_n({\mathbb C})$, the $n\times n$ complex
matrix algebra. For any non-identity permutation $\pi$ of
$(1,2,\ldots ,n)$, let $\Phi_{n,t,\pi}: M_n({\mathbb C}) \rightarrow
M_n({\mathbb C})$ be a linear map defined by
$$\Phi_{n,t,\pi}(A)=(n-t)\sum_{i=1}^nE_{ii}AE_{ii}+t\sum_{i=1}^nE_{i,\pi(i)}AE_{i,\pi(i)}^\dag-A,\eqno(1.1)$$
where   $0\leq t\leq n$, $E_{ij}$s are the matrix units, that is,
$E_{ij}$ is the matrix with $(i,j)$-entry 1 and other entries 0, and
$\pi$ is a non-identity permutation of $(1,2,\cdots,n)$.

According to \cite[Proposition 6.2]{HLPQS}, $\Phi_{n,t,\pi}$ is a
NCP positive map  if and only if $0< t\leq \frac{n}{l(\pi)}$. Hence
$W_{n,t,\pi}=W_{\Phi_{n,t,\pi}}$ induced by $\Phi_{n,t,\pi}$ is an
EW if and only if $0<t\leq \frac{n}{l(\pi)}$. It is also shown that
$\Phi_{n,1,\pi}$ is a decomposable NCP positive map  if $\pi$ is a
non-identity permutation of $(1,2,\ldots,n)$ with $\pi^2=$id
\cite[Proposition 7.2]{HLPQS}. This result implies that
$W_{n,1,\pi}$ is a decomposable EW if $\pi^2={\rm id}$.

In this paper, we will first prove in Section 2 that the condition
$\pi^2={\rm id}$ is in fact a necessary and sufficient condition for
the decomposability of entanglement witnesses $W_{n,t,\pi}$ for any
$0< t\leq \frac{n}{l(\pi)}$ (Theorem 2.1). Thus, we obtain a new and
large class of indecomposable entanglement witnesses $W_{n,t,\pi}$
constructed from any permutations $\pi$ with $\pi^2\not={\rm id}$.
To check the indecomposability of $W_{n,t,\pi}$ where
$\pi^2\not={\rm id}$, we construct some new bound entangled states
which can be detected by $W_{n,t,\pi}$. Section 3 is devoted to
comparing our EWs $W_{n,t,\pi}$ with other separability criteria and
show that there are  entangled states that can be detected by
$W_{n,t,\pi}$ but can not be detected by PPT criterion, realignment
criterion and an inequality criterion that even stronger than the
realignment criterion. In Section 4, a short conclusion is given.

\section{Necessary and sufficient condition for  $W_{n,t,\pi}$ to be indecomposable}

In this section we discuss the question: when  $W_{n,t,\pi}$ is
indecomposable? The following is our main result.

{\bf Theorem 2.1.} {\it Let $\pi$ be a non-identity permutation
 of $(1,2,\ldots, n)$ with   $n\geq 2$, and $0<t\leq \frac{n}{l(\pi)}$. The
entanglement witness $W_{n,t,\pi}$ is  indecomposable  if and only
if $\pi^2\not={\rm id}$.}

We need a simple lemma, which is a slight generalization of
\cite[Proposition 2.6]{QH1}.

{\bf Lemma 2.2.}  {\it Let
$$B_{(t_1,t_2\cdots ,t_n)}=\left(\begin{array}{ccccc} t_1
&-1&-1&\cdots &-1\\
-1 &t_2&-1&\cdots &-1 \\
\vdots &\vdots &\vdots&\ddots &\vdots\\
 -1 &-1&-1&\cdots &t_n
 \end{array}\right)\in M_n({\mathbb C}).$$ If  $0\leq t_i\leq n-1$ for each $i=1,2,
\cdots, n$ and there exists at least one $i_0\in\{1,2, \cdots, n\}$
such that $t_{i_0}<n-1$, then $B_{(t_1,t_2\cdots ,t_n)}\ngeq 0$.}

{\bf Proof.} Let
$|\psi_+\rangle=\frac{1}{\sqrt{n}}\sum_{i=1}^n|i\rangle$. Then
$$B_{(t_1,t_2\cdots ,t_n)}=n(I_n-|\psi_+\rangle\langle\psi_+|)-{\rm diag}(n-1-t_1,n-1-t_2,\cdots,n-1-t_n).$$
Take $\{|\psi_+\rangle,|\phi_1\rangle,\cdots,|\phi_{n-1}\rangle\}$
as another orthonormal basis of ${\mathbb C}^n$. We have
$|i_0\rangle=\alpha|\psi_+\rangle+\sum_{i=1}^{n-1}\alpha_i|\phi_i\rangle$
for scalars $\alpha,\alpha_1,\cdots,\alpha_{n-1}$. It is easily seen
that $\alpha=\langle\psi_+|i_0\rangle=\frac{1}{\sqrt{n}}$. So
$|i_0\rangle=\frac{1}{\sqrt{n}}|\psi_+\rangle+\sum_{i=1}^{n-1}\alpha_i|\phi_i\rangle$,
and hence $|i_0\rangle\langle
i_0|=\frac{1}{n}|\psi_+\rangle\langle\psi_+|+A$, where $A=
\frac{1}{\sqrt{n}}\sum_{i=1}^{n-1}(|\psi_+\rangle\langle\phi_i|+|\phi_i\rangle\langle\psi_+|)
+\sum_{i,j=1}^{n-1}\alpha_i\bar{\alpha}_j|\phi_i\rangle\langle\phi_j|$.
Thus, one obtains
$$\begin{array}{rl}B_{(t_1,t_2\cdots ,t_n)}\leq&
n(I_n-|\psi_+\rangle\langle\psi_+|)-(n-1-t_{i_0})|i_0\rangle\langle
i_0|\\
=&n(I_n-|\psi_+\rangle\langle\psi_+|)-\frac{n-1-t_{i_0}}{n}|\psi_+\rangle\langle
\psi_+|-(n-1-t_{i_0})A=B.\end{array}$$ Note that, under the space
decomposition ${\mathbb C}^n=[|\psi_+\rangle]\oplus
[|\psi_+\rangle]^\perp$,
$$B=\left(\begin{array}{cc} -\frac{n-1-t_{i_0}}{n}&B_{12}\\
B_{21}&B_{22}\end{array}\right),$$which is not positive
semi-definite obviously. It follows that $B_{(t_1,t_2\cdots
,t_n)}\ngeq 0$. \hfill$\Box$

{\bf Proof of Theorem 2.1.} To check the ``only if" part, assume
$\pi^2={\rm id}$. Let $F$ be the set of fixed points of $\pi$. Since
$\Phi_{n,t,\pi}(E_{ii})=(n-t-1)E_{ii}+tE_{\pi(i),\pi(i)}$ and
$\Phi_{n,t,\pi}(E_{ij})=-E_{ij}$ ($i\not=j$), we have
$$\begin{array}{rl}W_{{n,t,\pi}}=&\sum_{i=1}^n(n-1-t)E_{ii}\otimes E_{ii}+
\sum_{i=1}^ntE_{\pi(i),\pi(i)}\otimes E_{ii}-\sum_{i\not=j}E_{ij}\otimes E_{ij}\\
=&\sum_{i\in F}(n-1)E_{ii}\otimes E_{ii}+\sum_{i\not\in
F}(n-1-t)E_{ii}\otimes E_{ii}\\
&-\sum_{i\neq j; \pi(i)\neq j}E_{ij}\otimes E_{ij}+\sum_{i\not\in
F}tE_{\pi(i),\pi(i)}\otimes E_{ii}-\sum_{i\not\in
F}E_{i,\pi(i)}\otimes E_{i,\pi(i)}.\end{array}$$ Let
$$\begin{array}{rl}Q_1=&\sum_{i\in F}(n-1)E_{ii}\otimes E_{ii}+\sum_{i\not\in
F}(n-1-t)E_{ii}\otimes E_{ii}\\
&-\sum_{i\neq j; \pi(i)\neq j}E_{ij}\otimes E_{ij}-\sum_{i\not\in
F}(1-t)E_{i,\pi(i)}\otimes E_{i,\pi(i)}\end{array}$$and
$$Q_2=\sum_{i\not\in
F}tE_{\pi(i),\pi(i)}\otimes E_{ii}-\sum_{i\not\in
F}tE_{i,\pi(i)}\otimes E_{i,\pi(i)}.$$ Since $\pi^2={\rm id}$, the
cardinal number of $F^c$ must be even. Thus we have
$$Q_2=\sum_{i< \pi(i)}(tE_{\pi(i),\pi(i)}\otimes E_{ii}+tE_{ii}\otimes E_{\pi(i),\pi(i)}
-tE_{i,\pi(i)}\otimes E_{i,\pi(i)}-tE_{\pi(i),i}\otimes
E_{\pi(i),i}).$$ As
$$Q_2^{{\rm T}_2}=\sum_{i< \pi(i)}(tE_{\pi(i),\pi(i)}\otimes E_{ii}+tE_{ii}\otimes E_{\pi(i),\pi(i)}
-tE_{i,\pi(i)}\otimes E_{\pi(i),i}-tE_{\pi(i),i}\otimes
E_{i,\pi(i)})\geq 0,$$ we see that $Q_2$ is PPT. Observe that
$Q_1\cong B\oplus 0\not=0$, where $B=(b_{ij})\in M_n({\mathbb C})$
is a Hermitian matrix satisfying $b_{ii}=n-1$ or $n-1-t$,
$b_{ij}=t-1$ or $-1$ so that $\sum_{j=1}^n b_{ij}=0$ for each $i$.
It is easily seen from the strictly diagonal dominance theorem (Ref.
\cite[Theorem 6.1.10]{Horn}) that $B$ is semi-definite. So $Q_1\not=
0$ is positive semi-definite. Hence $W_{n,t,\pi}=Q_1+Q_2$ is
decomposable, completing the proof for the ``only if" part.

\if false If $1\leq t\leq\frac{n}{l(\pi)}$, then let
$$Q_1=\sum_{i\in F}(n-1)E_{ii}\otimes E_{ii}+\sum_{i\not\in
F}(n-1-t)E_{ii}\otimes E_{ii}-\sum_{i\neq j; \pi(i)\neq
j}E_{ij}\otimes E_{ij}$$and
$$Q_2=\sum_{i\not\in
F}tE_{\pi(i),\pi(i)}\otimes E_{ii}-\sum_{i\not\in
F}E_{i,\pi(i)}\otimes E_{i,\pi(i)}.$$ Since $\pi^2={\rm id}$, the
cardinal number of $F^c$ must be even. Thus we have
$$Q_2=\sum_{i< \pi(i)}(tE_{\pi(i),\pi(i)}\otimes E_{ii}+tE_{ii}\otimes E_{\pi(i),\pi(i)}
-E_{i,\pi(i)}\otimes E_{i,\pi(i)}-E_{\pi(i),i}\otimes
E_{\pi(i),i}).$$ As
$$Q_2^{{\rm T}_2}=\sum_{i< \pi(i)}(tE_{\pi(i),\pi(i)}\otimes E_{ii}+tE_{ii}\otimes E_{\pi(i),\pi(i)}
-E_{i,\pi(i)}\otimes E_{\pi(i),i}-E_{\pi(i),i}\otimes
E_{i,\pi(i)})\geq 0,$$ we see that $Q_2$ is PPT (positive partial
transpose).

Observe that $Q_1\cong A\oplus 0$, where $A=(a_{ij})\in M_n({\mathbb
C})$ is a Hermitian matrix satisfying $a_{ii}=n-2$ or $n-1$,
$a_{ij}=0$ or $-1$ so that $\sum_{j=1}^n a_{ij}=0$ for any $i$. It
is easily seen from the strictly diagonal dominance theorem  (Ref.
\cite[Theorem 6.1.10]{Horn}) that $A$ is semi-definite. So $Q_1\not=
0$ is positive semi-definite. Hence $W_{n,t,\pi}=Q_1+Q_2$ is
decomposable if $1\leq t\leq\frac{n}{l(\pi)}$.

If $0<t<1$, then let
$$\begin{array}{rl}Q_1=&\sum_{i\in F}(n-1)E_{ii}\otimes E_{ii}+\sum_{i\not\in
F}(n-1-t)E_{ii}\otimes E_{ii}\\
&-\sum_{i\neq j; \pi(i)\neq j}E_{ij}\otimes E_{ij}-\sum_{i\not\in
F}(1-t)E_{i,\pi(i)}\otimes E_{i,\pi(i)}\end{array}$$and
$$Q_2=\sum_{i\not\in
F}tE_{\pi(i),\pi(i)}\otimes E_{ii}-\sum_{i\not\in
F}tE_{i,\pi(i)}\otimes E_{i,\pi(i)}.$$ It is clear that $Q_2$ is
PPT. Also observe that $Q_1\cong B\oplus 0\not=0$, where
$B=(b_{ij})\in M_n({\mathbb C})$ is a Hermitian matrix satisfying
$b_{ii}=n-2$ or $n-1-t$, $b_{ij}=t-1$ or $-1$ so that $\sum_{j=1}^n
b_{ij}=0$ for each $i$. So $B$ is semi-definite and $Q_1\geq 0$.
Again, we get $W_{n,t,\pi}$ is decomposable for any $0<t<1$,
completing the proof for the `` if" part.\fi

Next we check the ``if" part, that is, we need  to show that
$\pi^2\not={\rm id}$ implies that $W_{n,t,\pi}$ is indecomposable.
Write $\pi=(\pi_1)(\pi_2)\cdots(\pi_k)$ with $\pi_s(F_s)=F_s$ and
$l(\pi_s)=l_s$, $s=1,2,\cdots,k$.  It is clear that $\pi^2\not={\rm
id}$ if and only if $l=l(\pi)\geq 3$.

If $l=n$, then $F_1=\{1,2,\cdots,n\}$ and $\pi$ is a cyclic
permutation. By \cite{QH1,QH}, we know that $W_{n,t,\pi}$ is
indecomposable.

Now    assume $3\leq l<n$. Without loss of generality, assume
$l=l_1\geq l_2\geq \cdots\geq l_k$ and $l_{m+1}=\cdots=l_k=1$ for
$1\leq m\leq k$.

Let $|\omega\rangle=\frac{1}{\sqrt{n}}\sum_{i=1}^n|i\rangle$. Define
$\rho_0=|\omega\rangle\langle\omega|$,
$$\rho_{sj}=\frac{1}{l_s}\sum_{i\in F_s}|i\rangle\langle i|\otimes|\pi_s^j(i)\rangle\langle\pi_s^j(i)|,\ j=1,2,\cdots,l_s-1;\ s=1,2,\cdots,m$$
and
$$\tilde{\rho}=\frac{1}{\sum_{s=1}^ml_s(n-l_s)}\sum_{s=1}^m\sum_{i\in F_s;j\not\in F_s}|j\rangle\langle j|\otimes |i\rangle\langle i|.$$
Let
$$\rho=q_0\rho_0+\sum_{s=1}^m\sum_{j=1}^{l_s-1}q_{sj}\rho_{sj}+\tilde{q}\tilde{\rho}\ \ \ {\rm with}\ \ \ q_0+\sum_{s=1}^m\sum_{j=1}^{l_s-1}q_{sj}+\tilde{q}=1.$$ For
such $\rho$, it can be checked that
$$\begin{array}{rl}&(\Phi_{n,t,\pi}\otimes I)(\rho)\\
\cong &A\oplus
(\frac{n-1-t}{l_1}q_{1,l_1-1}+\frac{t}{l_1}q_{1,l_1-2})I_{l_1}\oplus\cdots
\oplus(\frac{n-1-t}{l_1}q_{1,2}+\frac{t}{l_1}q_{1,1})I_{l_1}\oplus(\frac{n-1-t}{l_1}q_{1,1}+\frac{t}{n}q_{0})I_{l_1}\\
&\oplus\cdots\cdots\\&\oplus
(\frac{n-1-t}{l_k}q_{k,l_k-1}+\frac{t}{l_k}q_{k,l_k-2})I_{l_k}\oplus\cdots
\oplus(\frac{n-1-t}{l_k}q_{k,2}+\frac{t}{l_k}q_{k,1})I_{l_k}\oplus(\frac{n-1-t}{l_k}q_{k,1}+\frac{t}{n}q_{0})I_{l_k}\\
&\oplus
\frac{\tilde{q}}{\sum_{s=1}^kl_s(n-l_s)}I_{\sum_{s=1}^kl_s(n-l_s)},\end{array}$$
where
$$A={\small\left(\begin{array}{cccccccccccccc}
t_1&-\frac{q_0}{n}&\cdots&-\frac{q_0}{n}&-\frac{q_0}{n}&\cdots&-\frac{q_0}{n}&\cdots&-\frac{q_0}{n}&\cdots&-\frac{q_0}{n}&-\frac{q_0}{n}&\cdots&-\frac{q_0}{n}\\
-\frac{q_0}{n}&t_1&\cdots&-\frac{q_0}{n}&-\frac{q_0}{n}&\cdots&-\frac{q_0}{n}&\cdots&-\frac{q_0}{n}&\cdots&-\frac{q_0}{n}&-\frac{q_0}{n}&\cdots&-\frac{q_0}{n}\\
\vdots&\vdots&\ddots&\vdots&\vdots&\vdots&\vdots&\vdots&\vdots&\vdots&\vdots&\vdots&\vdots&\vdots\\
-\frac{q_0}{n}&-\frac{q_0}{n}&\cdots&t_1&-\frac{q_0}{n}&\cdots&-\frac{q_0}{n}&\cdots&-\frac{q_0}{n}&\cdots&-\frac{q_0}{n}&-\frac{q_0}{n}&\cdots&-\frac{q_0}{n}\\
-\frac{q_0}{n}&-\frac{q_0}{n}&\cdots&-\frac{q_0}{n}&t_2&\cdots&-\frac{q_0}{n}&\cdots&-\frac{q_0}{n}&\cdots&-\frac{q_0}{n}&-\frac{q_0}{n}&\cdots&-\frac{q_0}{n}\\
\vdots&\vdots&\vdots&\vdots&\vdots&\ddots&\vdots&\vdots&\vdots&\vdots&\vdots&\vdots&\vdots&\vdots\\
-\frac{q_0}{n}&-\frac{q_0}{n}&-\frac{q_0}{n}&\cdots&-\frac{q_0}{n}&\cdots&t_2&\cdots&-\frac{q_0}{n}&\cdots&-\frac{q_0}{n}&-\frac{q_0}{n}&\cdots&-\frac{q_0}{n}\\
\vdots&\vdots&\vdots&\vdots&\vdots&\vdots&\vdots&\ddots&\vdots&\vdots&\vdots&\vdots&\vdots&\vdots\\
-\frac{q_0}{n}&-\frac{q_0}{n}&-\frac{q_0}{n}&\cdots&-\frac{q_0}{n}&\cdots&-\frac{q_0}{n}&\cdots&t_m&\cdots&-\frac{q_0}{n}&-\frac{q_0}{n}&\cdots&-\frac{q_0}{n}\\
\vdots&\vdots&\vdots&\vdots&\vdots&\vdots&\vdots&\vdots&\vdots&\ddots&\vdots&\vdots&\vdots&\vdots\\
-\frac{q_0}{n}&-\frac{q_0}{n}&-\frac{q_0}{n}&\cdots&-\frac{q_0}{n}&\cdots&-\frac{q_0}{n}&\cdots&-\frac{q_0}{n}&\cdots&t_m&-\frac{q_0}{n}&\cdots&-\frac{q_0}{n}\\
-\frac{q_0}{n}&-\frac{q_0}{n}&-\frac{q_0}{n}&\cdots&-\frac{q_0}{n}&\cdots&-\frac{q_0}{n}&\cdots&-\frac{q_0}{n}&\cdots&-\frac{q_0}{n}&\frac{(n-1)q_0}{n}&\cdots&-\frac{q_0}{n}\\
\vdots&\vdots&\vdots&\vdots&\vdots&\vdots&\vdots&\vdots&\vdots&\vdots&\vdots&\vdots&\ddots&\vdots\\
-\frac{q_0}{n}&-\frac{q_0}{n}&-\frac{q_0}{n}&\cdots&-\frac{q_0}{n}&\cdots&-\frac{q_0}{n}&\cdots&-\frac{q_0}{n}&\cdots&-\frac{q_0}{n}&-\frac{q_0}{n}&\cdots&\frac{(n-1)q_0}{n}\end{array}\right)}
$$
with $t_i=\frac{n-1-t}{n}q_0+\frac{t}{l_i}q_{i,l_i-1}$,
$i=1,2,\cdots,m$. Thus, by Lemma 2.2, we get that $A\ngeq 0$ if
$\frac{n-1-t}{n}q_0+\frac{t}{l_i}q_{i,l_i-1}\leq \frac{n-1}{n}q_0$
for all $i=1,2,\cdots,m$ and there exists at least one $i_0$ such
that $\frac{n-1-t}{n}q_0+\frac{t}{l_{i_0}}q_{i_0,l_{i_0}-1}<
\frac{n-1}{n}q_0$. It follows that, if
$$q_{i,l_i-1}\leq \frac{l_i}{n}q_0,\ i=1,2,\cdots,m\ {\rm and \ at \ least\ one
}\ i_0\ {\rm such\ that}\
q_{i_0,l_{i_0}-1}<\frac{l_{i_0}}{n}q_0,\eqno(2.1)$$ then
$(\Phi_{n,t,\pi}\otimes I)(\rho)$ is not positive, and hence, the
state $\rho$ is entangled (the positive map criterion in \cite{Hor,
H}).

Note that $\rho$ is PPT if and only if the following two conditions
hold:
$$q_{s,l_s-i}q_{s,i}\geq \frac{l_s^2}{n^2}q_0^2, \ \ i=1,2,\cdots,l_s-1;\ \ s=1,2,\cdots,m\eqno(2.2)$$
and
$$\tilde{q}\geq \frac{\sum_{s=1}^ml_s(n-l_s)}{n}q_0.\eqno(2.3)$$
Moreover,  we can choose $q_0$, $q_{sj}$ and $\tilde{q}$ so that
Eqs.(2.1)-(2.3) hold simultaneously. For example, take
$\tilde{q}=\frac{\sum_{s=1}^ml_s(n-l_s)}{n}q_0$, and for
$i\in\{1,2,\cdots,m \}$, if $l_i$ is even, take
$$q_{i,1}=q_{i,2}=\cdots=q_{i,\frac{l_i}{2}-1}=q_0,\ \ q_{i,\frac{l_i}{2}}=\frac{l_i}{n},\ \ q_{i,\frac{l_i}{2}+1}=\cdots=q_{i,l_i-1}=\frac{l_i^2}{n^2}q_0;$$
if $l_i$ is odd, take
$$q_{i,1}=q_{i,2}=\cdots=q_{i,\frac{l_i-1}{2}}=q_0,\ \ q_{i,\frac{l_i-1}{2}+1}=\cdots=q_{i,l_i-1}=\frac{l_i^2}{n^2}q_0.$$
Such  $q_0$, $q_{sj}$ ($s=1,2,\cdots,m$, $j=1,2,\cdots,l_s-1$) and
$\tilde{q}$ satisfy Eqs.(2.1)-(2.3). It follows that $\rho$ is PPT
entangled which can be recognized by $\Phi_{n,t,\pi}$. Hence,
$\Phi_{n,t,\pi}$ is not decomposable, and consequently,
$W_{n,t,\pi}$ is indecomposable.

This completes the proof of  Theorem 2.1. \hfill$\Box$

To illustrate the structure of  the bounded entangled states
constructed in the proof of Theorem 2.1 to show that $W_{n,t,\pi}$
is indecomposable whenever $\pi^2\not=$id, we give two examples in
cases $n=4$ and $n=5$.

{\bf Example 2.3.} Let $\{|i\rangle\}_{i=1}^4$ be any orthonormal
basis of ${\mathbb C}^4$. Let $\pi$ be the permutation  of
$(1,2,3,4)$ defined by   $\pi(1)=2$, $\pi(2)=3$, $\pi(3)=1$ and
$\pi(4)=4$. Then $\pi^2\not=$id and $l=l(\pi)=3$. For such $\pi$,
the state $\rho$ in the proof of Theorem 2.1 is constructed as
follows.

Let $|\omega\rangle=\frac{1}{2}\sum_{i=1}^4|i\rangle.$ Define
$\rho_0=|\omega\rangle\langle\omega|$ and
$\rho_{11}=\frac{1}{3}(|1\rangle\langle 1|\otimes |2\rangle\langle
2|+|2\rangle\langle 2|\otimes |3\rangle\langle 3|+|3\rangle\langle
3|\otimes |1\rangle\langle 1|)$,
$\rho_{12}=\frac{1}{3}(|1\rangle\langle 1|\otimes |3\rangle\langle
3|+|2\rangle\langle 2|\otimes |1\rangle\langle 1|+|3\rangle\langle
3|\otimes |2\rangle\langle 2|)$ and
$\tilde{{\rho}}=\frac{1}{6}(\sum_{i=1}^3|i\rangle\langle
i|\otimes|4\rangle\langle 4|+\sum_{j=1}^3|4\rangle\langle
4|\otimes|j\rangle\langle j|)$. Let
$\rho=q_0\rho_0+q_{11}\rho_{11}+q_{12}\rho_{12}+\tilde{q}\tilde{\rho}$,
where $q_0,q_{11},q_{12},\tilde{q}\geq 0$ and
$q_0+q_{11}+q_{12}+\tilde{q}=1$.

For such $\rho$, it is easily checked that
$$(\Phi_{4,t,\pi}\otimes I)(\rho)
\cong A\oplus ( \frac{3-t}{4}q_{12}+\frac{t}{3}q_{11})I_3\oplus
(\frac{3-t}{3}q_{11}+\frac{t}{4}q_0)I_3\oplus \frac{1}{2}q_4I_6,$$
where
$$A={\left(\begin{array}{cccc}
\frac{3-t}{4}q_0+\frac{t}{3}q_{12}&-\frac{q_0}{4}&-\frac{q_0}{4}&-\frac{q_0}{4}\\
-\frac{q_0}{4}&\frac{3-t}{4}q_0+\frac{t}{3}q_{12}&-\frac{q_0}{4}&-\frac{q_0}{4}\\
-\frac{q_0}{4}&-\frac{q_0}{4}&\frac{3-t}{4}q_0+\frac{t}{3}q_{12}&-\frac{q_0}{4}\\
-\frac{q_0}{4}&-\frac{q_0}{4}&-\frac{q_0}{4}&\frac{3}{4}q_0\end{array}\right)}.
$$
Thus, by Lemma 2.2, we get that $A\ngeq 0$ if
$q_{12}<\frac{3}{4}q_0$. So $(\Phi_{4,t,\pi}\otimes I)(\rho)$ is not
positive and $\rho$ is entangled if $q_{12}<\frac{3}{4}q_0$. Note
that $\rho$ is PPT if and only if $\tilde{q}\geq \frac{3}{2}q_0$ and
$q_{11}q_{12}\geq \frac{9}{16}q_0^2$. Take
$q_0=q_{11}=\frac{16}{65}$, $\tilde{q}=\frac{3}{2}q_0=\frac{24}{65}$
and $q_{12}=\frac{9}{16}q_0=\frac{9}{65}$. Then $\rho$ is PPT
entangled which can be recognized by $\Phi_{4,t,\pi}$. Hence,
$\Phi_{4,t,\pi}$ is not decomposable, and consequently,
$W_{4,t,\pi}$ is indecomposable.

{\bf Example 2.4.} Let $\{|i\rangle\}_{i=1}^5$ be any orthonormal
basis of ${\mathbb C}^5$. Let $\pi$ be the permutation  of
$(1,2,3,4,5)$ defined by   $\pi(1)=2$, $\pi(2)=1$, $\pi(3)=4$,
$\pi(4)=5$ and $\pi(5)=3$. Clearly, $\pi^2\not=$id and $l=l(\pi)=3$.
For such $\pi$, we construct $\rho$ as follows.

Let $|\omega\rangle=\frac{1}{\sqrt{5}}\sum_{i=1}^5|i\rangle.$ Define
$\rho_0=|\omega\rangle\langle\omega|$ and
$\rho_{11}=\frac{1}{2}(|1\rangle\langle 1|\otimes |2\rangle\langle
2|+|2\rangle\langle 2|\otimes |1\rangle\langle 1|)$,
$\rho_{21}=\frac{1}{3}(|3\rangle\langle 3|\otimes |4\rangle\langle
4|+|4\rangle\langle 4|\otimes |5\rangle\langle 5|+|5\rangle\langle
5|\otimes |3\rangle\langle 3|)$,
$\rho_{22}=\frac{1}{3}(|3\rangle\langle 3|\otimes |5\rangle\langle
5|+|4\rangle\langle 4|\otimes |3\rangle\langle 3|+|5\rangle\langle
5|\otimes |4\rangle\langle 4|)$ and
$\tilde{\rho}=\frac{1}{12}(\sum_{i=3}^5\sum_{j=1}^2|i\rangle\langle
i|\otimes|j\rangle\langle
j|+\sum_{j=3}^5\sum_{i=1}^2|i\rangle\langle
i|\otimes|j\rangle\langle j|)$. Let
$\rho=q_0\rho_0+q_{11}\rho_{11}+q_{21}\rho_{21}+q_{22}\rho_{22}+\tilde{q}\tilde{\rho}$,
where $q_0,q_{11},q_{21},q_{22},\tilde{q}\geq 0$ and
$q_0+q_{11}+q_{21}+q_{22}+\tilde{q}=1$. For such $\rho$, it is
easily checked that
$$\begin{array}{rl}&(\Phi_{5,t,\pi}\otimes I)(\rho)\\
\cong &A\oplus ( \frac{4-t}{2}q_{11}+\frac{t}{5}q_0)I_2\oplus (
\frac{4-t}{3}q_{22}+\frac{t}{3}q_2)I_3\oplus\frac{4-t}{3}q_{21}+\frac{t}{5}q_0)I_3\oplus
\frac{\tilde{q}}{3}I_{12},\end{array}$$ where
$$A={\small\left(\begin{array}{ccccc}
\frac{4-t}{5}q_0+\frac{t}{2}q_{11}&-\frac{q_0}{5}&-\frac{q_0}{5}&-\frac{q_0}{5}&-\frac{q_0}{5}\\
-\frac{q_0}{5}&\frac{4-t}{5}q_0+\frac{t}{2}q_{11}&-\frac{q_0}{5}&-\frac{q_0}{5}&-\frac{q_0}{5}\\
-\frac{q_0}{5}&-\frac{q_0}{5}&\frac{4-t}{5}q_0+\frac{t}{3}q_{22}&-\frac{q_0}{5}&-\frac{q_0}{5}\\
-\frac{q_0}{5}&-\frac{q_0}{5}&-\frac{q_0}{5}&\frac{4-t}{5}q_0+\frac{t}{3}q_{22}&-\frac{q_0}{5}\\
-\frac{q_0}{5}&-\frac{q_0}{5}&-\frac{q_0}{5}&-\frac{q_0}{5}&\frac{4-t}{5}q_0+\frac{t}{3}q_{22}\end{array}\right)}.
$$
Thus, by Lemma 2.2, we get that $A\ngeq 0$ if either
$q_{11}\leq\frac{2}{5}q_0$ and $q_{22}<\frac{3}{5}q_0$; or
$q_{11}<\frac{2}{5}q_0$ and $q_{22}\leq\frac{3}{5}q_0$. So
 $\rho$ is
entangled if  either $q_{11}\leq\frac{2}{5}q_0$ and
$q_{22}<\frac{3}{5}q_0$; or $q_{11}<\frac{2}{5}q_0$ and
$q_{22}\leq\frac{3}{5}q_0$. Note that $\rho$ is PPT if and only if
$q_{11}\geq \frac{2}{5}q_0$, $\tilde{q}\geq \frac{12}{5}q_0$ and
$q_{21}q_{22}\geq \frac{9}{25}q_0^2$. By taking
$q_0=q_{21}=\frac{25}{129}$, $q_{11}=\frac{2}{5}q_0=\frac{10}{129}$,
$q_{22}=\frac{9}{25}q_0=\frac{9}{129}$ and
$\tilde{q}=\frac{12}{5}q_0=\frac{60}{129}$, we get that $\rho$ is
PPT entangled which can be recognized by $\Phi_{5,t,\pi}$. Hence,
$\Phi_{5,t,\pi}$ is not decomposable, and consequently,
$W_{5,t,\pi}$ is indecomposable.

From Theorem 2.1, the following corollary is immediate.

{\bf Corollary 2.5.} {\it Let $\pi$ be a permutation of
$(1,2,\ldots,n)$ with $n\geq 2$ and $\pi\not={\rm id}$. Let $\Phi
_{n,t,\pi}:M_n({\mathbb C)}\rightarrow M_n({\mathbb C})$ be the
positive map defined in Eq.(1.1) with $0<t\leq \frac{n}{l(\pi)}$.
Then, $\Phi_{n,t,\pi}$ is decomposable if and only if $\pi^2={\rm
id}$.}

\section{Comparison with some other entanglement
criteria}

 The entanglement witnesses  $W_{n,t,\pi}$ constructed in this paper can detect some entangled states that cannot be detected by PPT
criterion, as demonstrated in Example 2.3 and Example 2.4. In this
section, we will show by examples that such entanglement witnesses
$W_{n,t,\pi}$ can also detect some entangled states that cannot be
detected by the realignment criterion.

{\bf Example 3.1.} Let us consider the state $\rho$ in Example 2.3.
Take $q_{11}=\tilde{q}=10q_0$ and $q_{12}=xq_0$ for $x\geq0$. Then
 $q_0=\frac{1}{x+21}$ and the state $\rho$ becomes $\rho_x$ as below:
$$\begin{array}{l}\rho_x=\\ {\small\left(\begin{array}{cccc|cccc|cccc|cccc}
\frac{q_0}{4}&0&0&0&0&\frac{q_0}{4}&0&0&0&0&\frac{q_0}{4}&0&0&0&0&\frac{q_0}{4}\\
0&\frac{xq_0}{3}&0&0&0&0&0&0&0&0&0&0&0&0&0&0\\
0&0&\frac{10q_0}{3}&0&0&0&0&0&0&0&0&0&0&0&0&0\\
0&0&0&\frac{10q_0}{6}&0&0&0&0&0&0&0&0&0&0&0&0\\ \hline
0&0&0&0&\frac{10q_0}{3}&0&0&0&0&0&0&0&0&0&0&0\\
\frac{q_0}{4}&0&0&0&0&\frac{q_0}{4}&0&0&0&0&\frac{q_0}{4}&0&0&0&0&\frac{q_0}{4}\\
0&0&0&0&0&0&\frac{xq_0}{3}&0&0&0&0&0&0&0&0&0\\
0&0&0&0&0&0&0&\frac{10q_0}{6}&0&0&0&0&0&0&0&0\\ \hline
0&0&0&0&0&0&0&0&\frac{xq_0}{3}&0&0&0&0&0&0&0\\
0&0&0&0&0&0&0&0&0&\frac{10q_0}{3}&0&0&0&0&0&0\\
\frac{q_0}{4}&0&0&0&0&\frac{q_0}{4}&0&0&0&0&\frac{q_0}{4}&0&0&0&0&\frac{q_0}{4}\\
0&0&0&0&0&0&0&0&0&0&0&\frac{10q_0}{6}&0&0&0&0\\ \hline
0&0&0&0&0&0&0&0&0&0&0&0&\frac{10q_0}{6}&0&0&0\\
0&0&0&0&0&0&0&0&0&0&0&0&0&\frac{10q_0}{6}&0&0\\
0&0&0&0&0&0&0&0&0&0&0&0&0&0&\frac{10q_0}{6}&0\\
\frac{q_0}{4}&0&0&0&0&\frac{q_0}{4}&0&0&0&0&\frac{q_0}{4}&0&0&0&0&\frac{q_0}{4}\end{array}\right)}.
\end{array}$$ By Theorem 2.1 and Example 2.3, we know that $\rho_x$ is
entangled which can be recognized by $W_{4,t,\pi}$ if
$x\in[0,\frac{3}{4})$ where $\pi: (1,2,3,4)\rightarrow (2,3,1,4)$.
However, by Example 2.3,   the entanglement in $\rho_x$ can be
detected by PPT criterion  only for $ x\in[0,\frac{9}{160})$.

\begin{figure}[h]
\begin{minipage}[t]{0.50\linewidth}
\setcaptionwidth{5in} \centering
{\includegraphics[width=\textwidth]{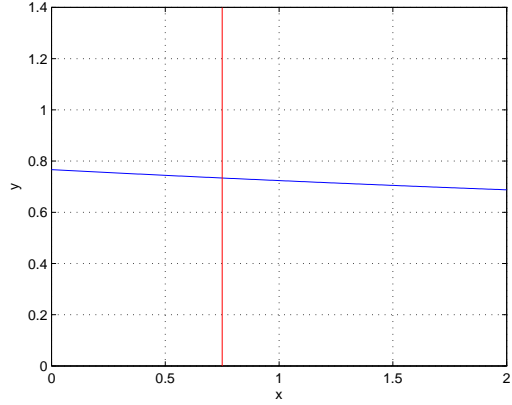} }\caption{ $y$-axis
denotes the value $\|R(\rho_x)\|_1$. Red and blue lines correspond
respectively to  the functions $x=\frac{3}{4}$ and
$y=\|R(\rho_x)\|_1$.}\label{1}
\end{minipage}
\end{figure}

Now, let us apply the realignment criterion (Ref. \cite{CW,GH2}) to
$\rho_x$. By a computation, for all $x\in[0,2]$, the trace norm of
the realignment $R(\rho_x)$ of $\rho_x$  is
$$\|R(\rho_x)\|_1=\frac{36+2\sqrt{16x^2-172x+1489}+\sqrt{y_+}+\sqrt{y_-}}{12(21+x)}<0.8<1,$$ where
$$\begin{array}{rl}  y_{\pm}=&8x^2+172x+2129\\ &\pm\sqrt{(16x^2+172x+1320)(172x+520)
+300(8x+92)^2+(8x^2+400)^2}\end{array}$$ (Ref. Figure \ref{1}).
 It follows that the entanglement in $\rho_x$ for
$x\in[0,\frac{3}{4})$ can be detected by $W_{4,t,\pi}$ but cannot be
distinguished by the realignment criterion.

In \cite{GH,ZZZG}, the authors proved that, if $\rho\in{\mathcal
S}(H_A\otimes H_B)$ is separable, then
$$\|R(\rho-\rho_A\otimes\rho_B)\|_{\rm Tr}\leq\sqrt{[1-{\rm
Tr}(\rho_A^2)][1-{\rm Tr}(\rho_B^2)]},\eqno(3.1)$$ where $\rho_A$
and $\rho_B$ are the reduced states with respect to subsystems $A$
and $B$, respectively. Thus, if $\rho$ breaks the inequality (3.1),
then $\rho$ is entangled. Furthermore, the inequality (3.1) provides
a stronger criterion than the realignment criterion.

Here, we will show by an example that our $W_{n,t,\pi}$ can also
detect some entangled states that cannot be detected by the
inequality (3.1).

\begin{figure}[h]
\begin{minipage}[t]{0.50\linewidth}
\setcaptionwidth{5in} \centering
{\includegraphics[width=\textwidth]{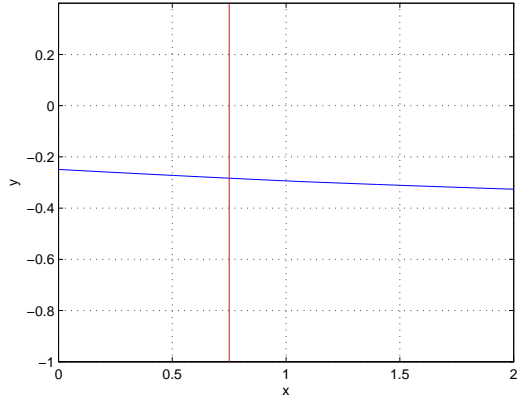} }\caption{$y$-axis
denotes the value of $\|R(\rho_x-\rho_{x,A}\otimes\rho_{x,B})\|_{\rm
Tr}-(1-{\rm Tr}(\rho_{x,A}^2))$. Red and blue lines correspond
respectively to the functions $x=\frac{3}{4}$ and
$y=f(x)$.}\label{2}
\end{minipage}
\end{figure}

{\bf Example 3.2.} Take $\rho_x$ and $W_{4,t,\pi}$ as in Example
3.1. Then we have
$$\rho_{x,A}=\rho_{x,B}={\left(\begin{array}{cccc}
\frac{4x+63}{12}q_0&0&0&0\\
0&\frac{4x+63}{12}q_0&0&0\\
0&0&\frac{4x+63}{12}q_0&0\\
0&0&0&\frac{21}{4}q_0\end{array}\right)}.$$ It is easily checked
that $$1-{\rm Tr}(\rho_{x,A}^2)=1-{\rm
Tr}(\rho_{x,B}^2)=\frac{32x^2+1512x+15876}{48(x+21)^2}$$ and thus
$$\sqrt{[1-{\rm
Tr}(\rho_A^2)][1-{\rm Tr}(\rho_B^2)]}=1-{\rm Tr}(\rho_{x,A}^2).$$
Since the realignment of $\rho_x-\rho_{x,A}\otimes\rho_{x,B}$ is
$$R(\rho_x-\rho_{x,A}\otimes\rho_{x,B})\cong \left(\begin{array}{cccc}
\frac{1}{4}q_0-u&\frac{x}{3}q_0-u&\frac{10}{3}q_0-u&\frac{10}{6}q_0-v\\
\frac{10}{3}q_0-u&\frac{1}{4}q_0-u&\frac{x}{3}q_0-u&\frac{10}{6}q_0-v\\
\frac{x}{3}q_0-u&\frac{10}{3}q_0-u&\frac{1}{4}q_0-u&\frac{10}{6}q_0-v\\
\frac{10}{6}q_0-v&\frac{10}{6}q_0-v&\frac{10}{6}q_0-v&\frac{1}{4}q_0-w\end{array}\right)\oplus\frac{1}{4}q_0I_{12},$$
where $u=(\frac{4x+63}{12})^2q_0^2$, $v=\frac{7(4x+63)}{16}q_0^2$
and $w=\frac{441}{16}q_0^2$, we have
$$\begin{array}{rl}\|R(\rho_x-\rho_{x,A}\otimes\rho_{x,B})\|_1=&2\sqrt{b-a}+\sqrt{\frac{2a+b+d+\sqrt{4a^2+4ab+(b-d)^2+12c^2-4ad}}{2}}\\
&+\sqrt{\frac{2a+b+d-\sqrt{4a^2+4ab+(b-d)^2+12c^2-4ad}}{2}}+3q_0,\end{array}$$
where
$$a=(\frac{1}{4}q_0-u)(\frac{x}{3}q_0-u)+(\frac{1}{4}q_0-u)(\frac{10}{3}q_0-u)+(\frac{x}{3}q_0-u)(\frac{10}{3}q_0-u)^2+(\frac{10}{6}q_0-v)^2,$$
$$b=(\frac{1}{4}q_0-u)^2+(\frac{x}{3}q_0-u)^2+(\frac{10}{3}q_0-u)^2+(\frac{10}{6}q_0-v)^2,$$
$$c=(\frac{1}{4}q_0+\frac{x}{3}q_0+\frac{10}{3}q_0-3u)(\frac{10}{6}q_0-v)+(\frac{10}{6}q_0-v)(\frac{1}{4}q_0-w)$$
and  $$d=3(\frac{10}{6}q_0-v)^2+(\frac{1}{4}q_0-w)^2.$$ By a
computation, we get
$$y=f(x)=\|R(\rho_x-\rho_{x,A}\otimes\rho_{x,B})\|_{\rm Tr}-(1-{\rm
Tr}(\rho_{x,A}^2))<-0.2<0$$ for $x\in[0,2]$ as shown by Figure
\ref{2}, which implies that the entanglement in $\rho_x$ for
$x\in[0,\frac{3}{4})$ cannot be distinguished by the inequality
(3.1).

\section{Conclusion}

By every non-identity permutation $\pi$ of $(1,2,\ldots , n)$ and
$0<t\leq \frac{n}{l(\pi)}$, where $l(\pi)$ is the length of $\pi$,
we can construct an entanglement witness $W_{n,t,\pi}$ for $n\otimes
n$ quantum system. $W_{n,t,\pi}$ is indecomposable if and only if $
\pi^2\not={\rm id}$. Thus a  class of  indecomposable entanglement
witnesses is obtained. Applying such witnesses, some new entangled
states, bounded entangled  states (that is, PPT entangled states)
are found. Several examples show that the entanglement witnesses
constructed in this paper can detect entanglement in some states
that cannot be detected by PPT criterion, the realignment criterion
and an inequality criterion stronger than the realignment criterion.

\end{document}